\documentclass[pre,reprint,showpacs,floatfix,amsmath,amssymb]{revtex4-1}

\usepackage{amsmath, amssymb, amsfonts, amsthm}
\usepackage{verbatim}
\usepackage{hyperref}
\usepackage{float} 
\usepackage[caption=false]{subfig}
\usepackage{graphicx,color}
\usepackage[colorinlistoftodos,prependcaption,textsize=small]{todonotes}

\begin{document}

\title{Translocation through conical pores: A direction-dependent process}

\author{Andri Sharma}
\email{ph17007@iisermohali.ac.in}

\affiliation{Department of Physical Sciences, Indian Institute of Science Education and Research (IISER) Mohali, Sector 81, S. A. S. Nagar 140306 Punjab, India}

\date{\today}% It is always \today, today,
             %  but any date may be explicitly specified

\begin{abstract}
The transport of biomolecules across a cell membrane is an important phenomena that plays a pivotal role in the functioning of biological cells. In this paper, we investigate such processes by modeling the translocation of polymers through a conical channel, directed from the wider opening to the narrow end of the conical channel. We use the molecular dynamics approach to study the problem. The effect of the different conical pore geometry and polymer lengths on translocation dynamics is determined from the behavior of the total translocation time, $\tau$, and  waiting time distributions, $w(s)$. The escape of polymer segments from the narrow end of the conical channel is tracked by studying their velocity profile ($v_{s}$). To demonstrate the asymmetric pore effects on the translocatin dynamics, we compare the translocation process from both the ends of the conical channel. We find striking differences in the translocation dynamics for both processes, which are in agreement with the experimental study. We have accounted the effect of various rigidity, and increasing length of a polymer chain, on both types of processes. The study can be used to find the transition from a directional dependent to a directional independent translocation process through a asymmetric channel.
\end{abstract}

%\keywords{Suggested keywords}%Use showkeys class option if keyword
                              %display desired
\maketitle

%\tableofcontents

\section{INTRODUCTION} 
The transport of biomolecules through bio-channels is a ubiquitous process and results in the emergence of various cell activities inside lipid membranes. Since the pioneering experiments on translocation by Bezrukov\cite{bezrukov1994counting} and Kasianowicz et al.\cite{kasianowicz1996characterization} several single-molecule experiments have been carried out. The objective of the experiments is to read the sequences of DNAs/RNAs by translocating the molecules across the membranes \cite{,akeson1999microsecond,meller2000rapid}.  The transport of macromolecules is involved in various biological and biotechnological phenomenons such as DNA packaging in capsid \cite{berndsen2014nonequilibrium}, gene exchange therapy, virus injection, etc \cite{lodish2008molecular}.   %\cite{palyulin2014polymer,branton2010potential,wanunu2012nanopores,howorka2009nanopore,keyser2011controlling,milchev2011single,bezrukov1994counting,kasianowicz1996characterization,sung1996polymer,park1998polymer,muthukumar1999polymer,muthukumar2003polymer,sakaue2007nonequilibrium,sakaue2010sucking,saito2012process,sarabadani2014iso,rowghanian2011force,ikonen2012unifying,ikonen2013influence,ikonen2012influence,bhattacharya2010out,bhattacharya2013translocation,de2010mapping,gauthier2008monte,gauthier2008monte2,huang2014conformations,luo2007influence,sarabadani2018theory,suhonen2018dynamics,katkar2018role}.
In solid state physics, artificial solid nanopores are fabricated to investigate the trapping and dynamics of single molecules  \cite{branton2008potential}, and this technique has led to a great advancement in polymer physics. Due to its vast range of applications, the transport phenomenon is one of the active areas of research for the last three decades in both computational and the experimental science  \cite{palyulin2014polymer}.

\begin{figure*}
\centering
{\includegraphics[scale=0.6]{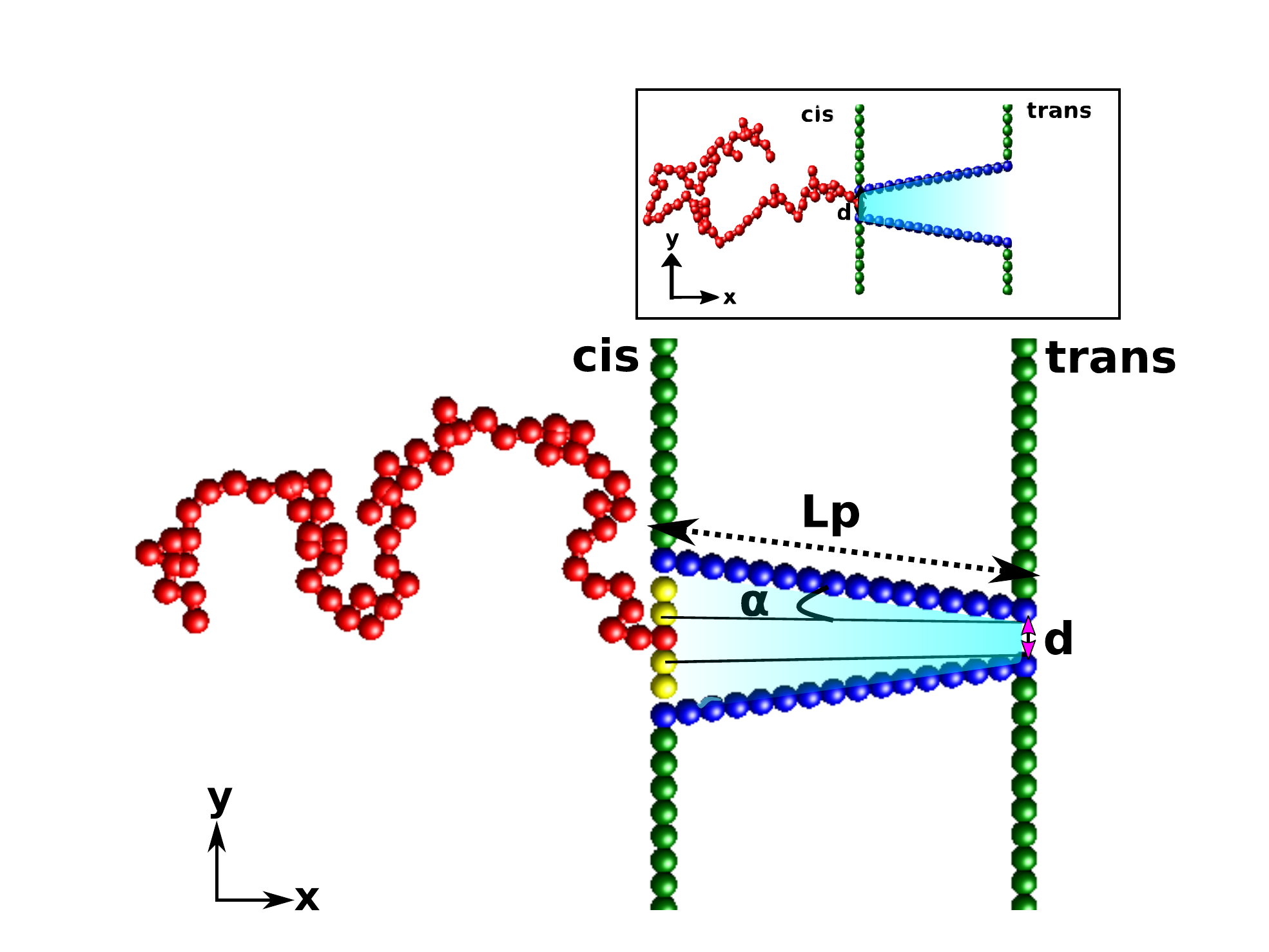}}
\caption{Schematic of an equilibrated flexible polymer sample for reverse translocation. The first bead of the chain is fixed at the entry of the conical pore, a few additional beads along with the first bead of the polymer are present at the pore entrance to restrict the entry of any polymer segments during the equilibration process. Inset shows the schematic of the equilibrated polymer chain for the forward translocation case, where the translocation takes place from the narrower end to the wider end of the conical channel. The color gradient in the channels represent the force gradient.}\label{reverselabel}
\end{figure*}

%A lot of experimental and computational work has been carried out to determine and verify the sequence of DNA through the translocation process 
Biological pores like Staphylococcus
aureus $\alpha$-hemolysin protein pores and MspA (Mycobacterium smegnatis Porin-A) pores are used to determine the sequence of a single  DNA and RNA molecule \cite{butler2006determination,butler2007ionic,schneider2012dna}. The sequences of DNA/RNA are differentiated at the narrow constriction of the pores from the current level detection method \cite{ashkenasy2005recognizing,stoddart2009single,vercoutere2001rapid,wong2010polymer,derrington2010nanopore,manrao2012reading}. Both MsPA and $\alpha$-hemolysin pores are similar in construction with a slight difference in their barrel structure. Whereas $\alpha$-hemolysin extended barrel is cylindrical in shape, MspA barrel is conical in shape \cite{schneider2012dna}. It is due to this conical shape of the MspA pore that the current was detected with only a few (3-4) nucleotides at the narrow constriction of the pore, whereas in $\alpha$-hemolysin pore, the detection required, at least, as many as 10 base pairs. Although the stability of level current is a challenge, these experiments paved the way to explore the interesting properties of biomolecules from sequencing through conical nanopores. Various synthetic nanopores are fabricated to study controlled transportation of DNA through cone-shaped artificial pores   \cite{li2004conical,asymmetric,zhou2017enhanced,liu2012voltage,chen2021dynamics}. This controlled transport of DNA molecules proved to be a success to detect various folded configurations of DNA inside conical nano capillaries \cite{thacker2012studying}.
In computational study, different types of channels are modeled so far to study the effect of pore-polymer interaction on the translocation processes \cite{nagarajan2020polyelectrolyte,mohan2010polymer}. But, of all the theoretical works carried out for extended pores, very few works have been performed on the study of translocation process via conical channels  
\cite{domanski2022simulation}. From the findings of previous works \cite{nikoofard2013directed,nikoofard2015flexible,sharma2022driven}, it is seen that the entropic drive due to the asymmetric structure of conical pores is non-monotonic as a function of cone apex angle ($\alpha$), and this non-monotonic effect is witnessed in the passage as well as the total translocation time ($\tau$). Due to its asymmetric shape, the translocation dynamics are completely dependent on the direction of conical pores\cite{asymmetric}. A detailed study is required to weigh the degree of differences of the translocation process from different directions of a conical channel.
In the present paper, we will investigate in details the translocation process from the wider openings of the conical vhannel and analyze the dependence of the translocation process on the direction of the conical pore. This work is the extension of our previous work for translocation through conical channels \cite{sharma2022driven}, where the effect of pore asymmetry was clearly reflected in the total translocation ($\tau$) and the waiting time distributions ($w(s)$) profile. Here, we tried to capture the transport of polymer from within the cell through the tube, by performing the translocation from the wider end (entrance) of the conical pore to the narrower exit. We call this process reverse translocation in contrast to our previous work \cite{sharma2022driven}, where we carried the translocation study from the narrow end of the conical channel to the wider end of the conical channel henceforth called forward translocation. It is seen that the direction of transport through symmetric channels does not affect the translocation properties\cite{gershow2007recapturing}. But as soon as this symmetry breaks down which is the case for the conical pores, the translocation properties tend to be direction-dependent for the same simulation parameters \cite{bell2017asymmetric,chen2021dynamics}. The nature of the driving force depends on the asymmetric design of the conical channels \cite{nikoofard2013directed}. In our studies, we have considered the effect of pore asymmetry on the nature of the driving force by choosing a specific form of force ($f_{x}$) which depends on channel variables $(\alpha,x)$, where $\alpha$ is the cone apex angle and x is the distance from the wider end(see Fig.~\ref{reverselabel}). We have established results that are supported by the experimental works \cite{bell2017asymmetric,chen2021dynamics}. We have performed Langevin Dynamics Simulation to study the problem in the weak force regime, so that the pore effect does not get washed out. Due to the asymmetry of the conical pore and the position-dependent external force, each bead will experience different velocities which result in the formation of different segment structures inside the pore. Therefore, the length and the coiling effects of the polymer near the wide entrance of the pore are studied by reading the escape bead velocity profile. We have compared the translocation properties for both the forward and the reverse translocation processes and have also taken the effect of rigidity to it. Furthermore, as mentioned above the validation of this work is in complete agreement with the experiment. This work is the extension of our previous work on translocation through conical channels \cite{sharma2022driven}. We will also discuss the limiting cases of this work.

The paper is organized as follows: In Section \ref{sec:model}, we define our model and simulation details. In Section \ref{sec:result}, we present our simulation results. We discuss our results and compare them with the previously published work. We summarise our results in Section \ref{sec:discussion}. 
%To see the effect of the coiling of polymer outside the pore and the resulting entropic force on the process, one needs to calculate the speed $v$ with which the polymer is translocating. If v is independent of the polymer length, then the primary cause of the polymer drag is due to the asymmetric pore. The force acting on the part of the polymer inside the pore is mainly due to the asymmetric shape of the pore. This force is taken into account in the form of x-dependent force $f_{x}${same in our previous paper}.      
%\figOne

\begin{figure}
\centering
{\includegraphics[width=0.5\textwidth]{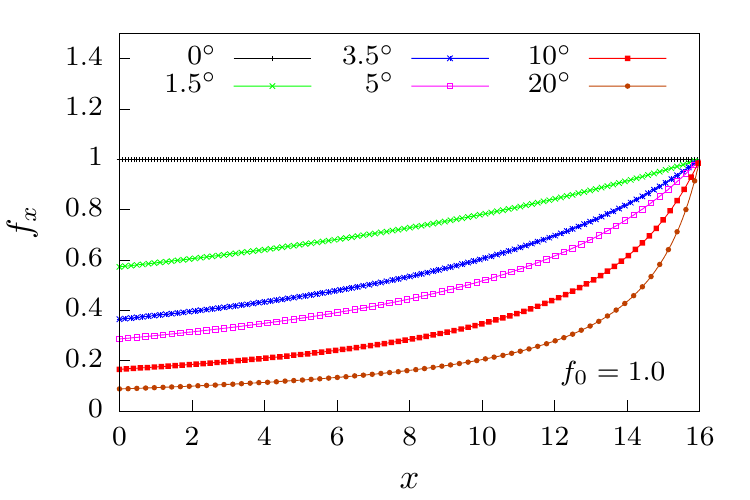}}
\caption{Variation of the magnitude of external force, $f_{x}$, along the conical channel for various cone angles ($\alpha$). For any non-zero $\alpha$ value, $f_{x}$ increases towards the narrow exit of the pore.}\label{forceplot}
\end{figure}

\section{MODEL AND METHOD} \label{sec:model}

%We consider 2D geometry for the polymer, wall, and channel system, in which the walls and channel are static. The linear homo polymer is modeled by a coarse-grained bead spring model and the length of the polymer, N, is 64 beads. 
We simulate our model in 2-dimensions. We model a linear homopolymer of length $N$ by a coarse-grained bead-spring model. The consecutive beads of the chain are connected  via harmonic potential,
\begin{equation}\label{eq:Ubond}
    U_{\textrm{bond}}=\frac{1}{2}K(r-r_0)^{2},
\end{equation}
where $K$ is the spring constant, and $r_0$ is the equilibrium separation between the consecutive monomers. The non-bonded beads interact via repulsive Lennard-Jones (LJ) potential, $U_{\textrm{bead}}(r)$. 
\begin{equation} \label{eq:Ubead}
U_{\textrm{bead}}(r)=
  \begin{cases}
  4 \epsilon \Bigg[ \left(\frac{\sigma}{r}\right)^{12}-\left(\frac{\sigma}{r}\right)^{6} \Bigg]+\epsilon,& r < r_{c}\\ 
  0,& r \geq r_{c}, 
  \end{cases}
\end{equation}
where $\epsilon$ is the depth of the regular LJ potential. The truncated and shifted LJ potential has a cut-off at  $r_{c}=2^{\frac{1}{6}}\sigma$, where $\sigma$ is the diameter of the bead.
%, where  $\sigma$ and $\epsilon$ are the distance and the energy, respectively. 
The static conical channel is constructed from $L_{p}=16$. There exists a long-range attractive LJ($U_{LJ}$) interaction between the polymer beads and the channel beads:
\begin{equation}\label{eq:ULJ}
U_{LJ}(r)=
  \begin{cases}
  4 \epsilon \Bigg[ \left(\frac{\sigma}{r}\right)^{12}-\left(\frac{\sigma}{r}\right)^{6} \Bigg],& r \leq R_{c} \\ 
  0,& \text{otherwise} 
  \end{cases}
\end{equation}
where $R_{c}=2.5\sigma$ is the cut-off distance. The motion of polymer beads is governed by the Langevin dynamics (LD) equation:
\begin{equation}
\label{eq:LD}
m \Ddot{\vec{r_{i}}}=-\eta\dot{\vec{r_{i}}}-\nabla\sum{(U_{LJ}+U_{bond})}+\vec{f_{ext}}+\zeta
\end{equation}
where $m$ and $r_{i}$ are, respectively, the mass and the position of $i$th bead of the polymer, $\eta$ is the friction coefficient of the fluid, and $\zeta$ is an uncorrelated random force which follows the fluctuation-dissipation relation,  $\langle \zeta_{i}(t)\zeta_{j}(t^{'}) \rangle = 2 K_{B}T\eta\delta_{ij}\delta(t-t^{'})$. The polymer experiences a force-gradient, $f_{x}$, along the $x$-direction of the channel:
\begin{equation}\label{fx}
    f_{x}=\frac{f_{0}d}{(d+2(16*tan\alpha))-(d+2(x*tan\alpha))}
\end{equation}

 $f_{0}$ is the magnitude of the force at the narrow opening of the conical channel, and $d$ is the diameter of the narrow end of the channel (see Fig.~\ref{forceplot}).

%The conical pore is designed in a way such that only the inner portion of the attractive cone with cone angle $2\alpha$ contributes to the asymmetric force field in the translocation process, and the role of the outer remaining portion $2(\pi-\alpha)$ is replaced by the solid repulsive wall on both, cis as well as trans side. \\

We set all the simulation parameters in reduced LJ units: $\epsilon$(energy unit), $\sigma$(the length unit), and m (mass). The dimensionless parameter $k_{B}T$, $\eta$ are set to be equal to 1. The strength of the force at the narrow end is taken to be $f_{0}= 0.2$ or stated otherwise. This value is found to be apt to capture the transport properties without washing out the role of the pore effect. Due to the wide opening at the entrance of the pore, there is no successful translocation in the absence of external force. All the simulations are carried out using LAMMPS software \cite{plimpton1995fast,thompson2022lammps}.
Before the translocation starts, the system is equilibrated using eq.~\eqref{eq:LD}. As can be seen in Fig.~\ref{reverselabel}, few additional yellow are constructed at the wider entrance of the conical pore to prevent the polymer segment from entering the channel during the equilibration process. Once the polymer attains the equilibrium configurations, the yellow beads are removed. Note that due to the wider opening at the entrance, multiple beads can enter the pore along with the first bead. In our study we consider averaging over only those samples for which the process begins with the entry of the first bead of the polymer chain, otherwise, the sample is discarded. The total translocation time, $\tau$, is the time lapse between the entrance of the first bead of the polymer chain into the channel from the cis side of the conical pore and the exit of the last bead of the polymer from the narrow end of the conical channel to its trans side. For each set of simulation parameters, the translocation time ($\tau$) is averaged over 1500-2000 independent samples.

\section{RESULTS AND DISCUSSION:} \label{sec:result}

\begin{figure}
\centering
\includegraphics[height=2.6in]{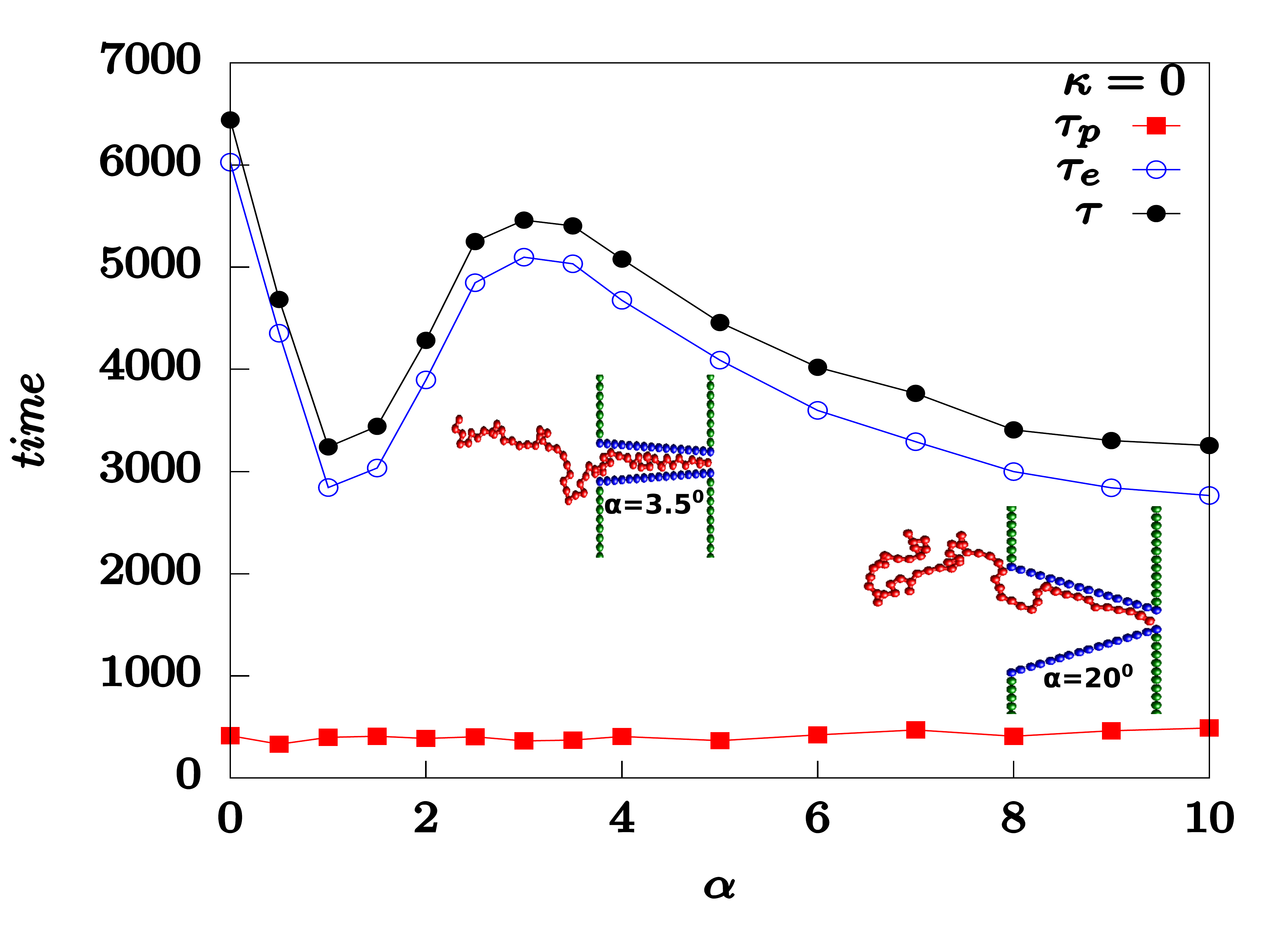}
\caption{Plot of total time  translocation ($\tau$) for flexible polymer($\kappa=0$) of length 64 beads along with the two time components: passage time ($\tau_{p}$) and escape time ($\tau_{e}$), s.t. $\tau=\tau_{p}+\tau_{e}$. Two simulation snapsorts are embedded in the plot to provide a visual scenario of the translocation process at lower and higher apex angle.}\label{reversetransK0}
\end{figure}

\begin{figure}
\includegraphics[width=0.5\textwidth]{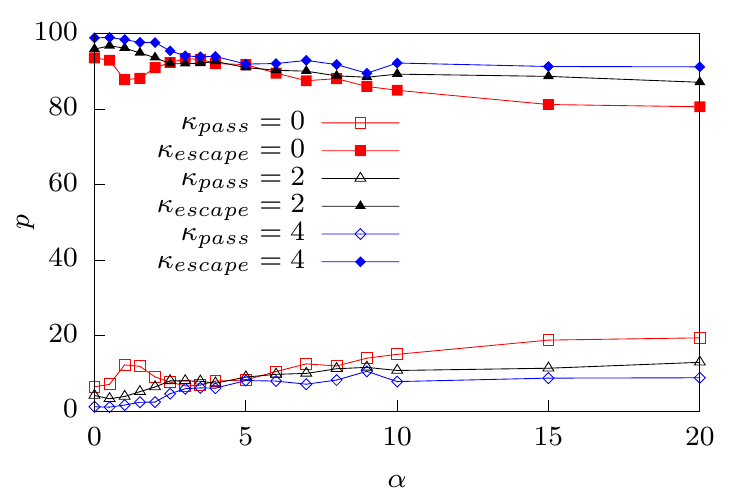}
\caption{Percentage contributions ($p$) from the discrete-time components, passage ($\tau_{p}$) and escape ($\tau_{e}$) times, to the total translocation time ($\tau$) for polymers of different rigidities. Here, $\kappa_{pass}$ and $\kappa_{escape}$ represent passage and escape time percentages from the respective polymer rigidities ($\kappa$).  }\label{reversepercent}
\end{figure}

\begin{figure*}
\includegraphics[width=\textwidth]{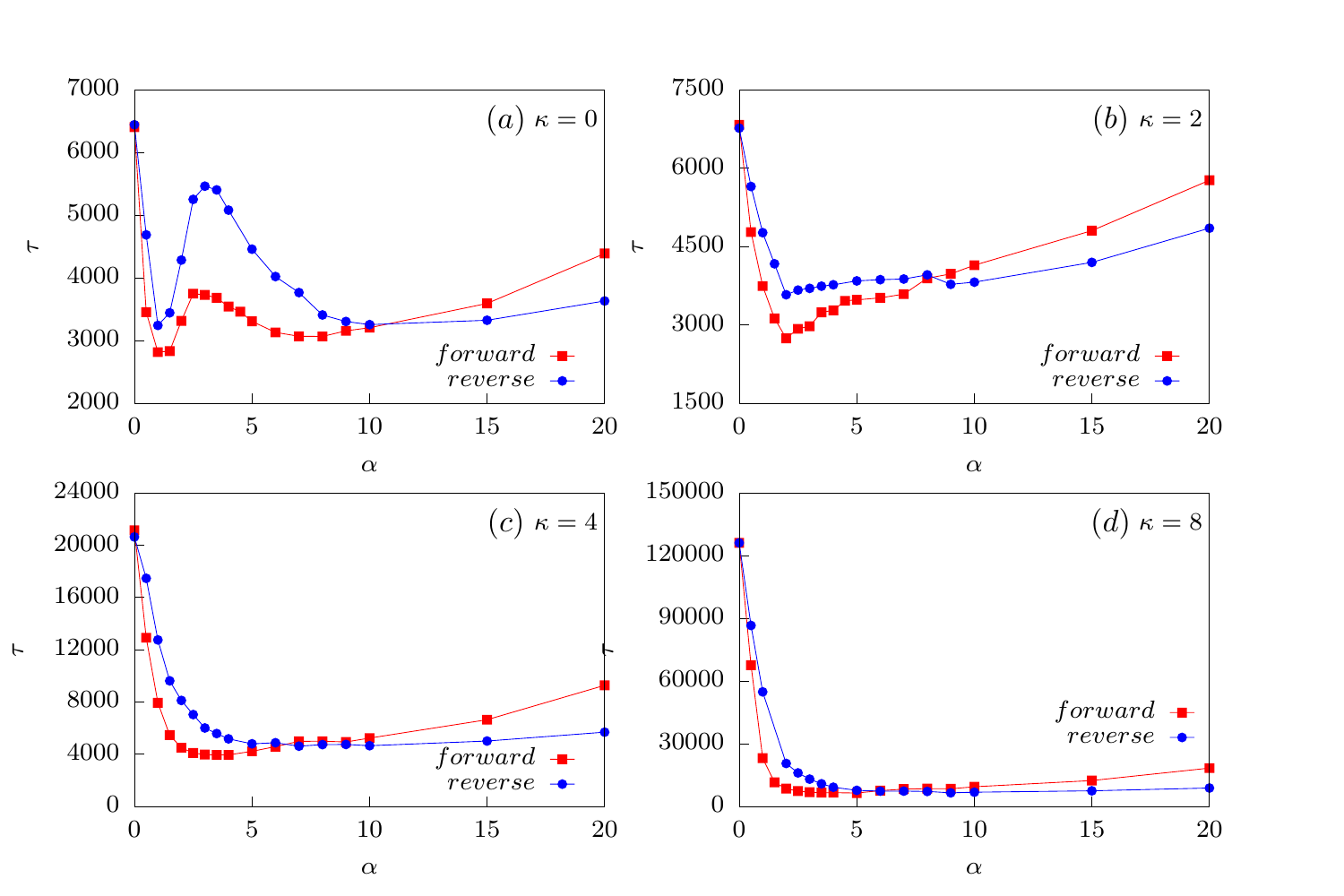}
\caption{ Comparison of forward (filled square) and reverse (filled circles) total translocation time ($\tau$) vs. cone apex angle ($\alpha$) at lower force value, $f_{0}=0.2$. Fig.(a-d) is plotted for $\kappa's=$0, 2, 4 and 8 respectively.}\label{forrevcompare}
\end{figure*}

\begin{comment}
\begin{figure*}
\includegraphics[height=5in]{transK0K2K4F0-2.pdf}
\caption{Total translocation time $\tau$ vs. half the apex angle of cone $\alpha$ for different flexibility of polymer with a chain length of $64$ beads. The plots in Fig.(a-c) are for three rigid polymer, $\kappa=$0, 2 and 4 in (a), (b) and (c) respectively. The three inter plots with different colours are: red for total translocation time $\tau$, blue one indicates the escape time, $\tau_{escape}$, and the olive is the filling time, $\tau_{passage}$, of the pore. And Fig.(d), is the $\%$ contributions from the two-time division of total time: escape time and passage time (same as filling time).  }\label{fig:reversek0k2k4F02}
\end{figure*}
\end{comment}

\begin{comment}
\begin{figure*}
\includegraphics[height=5in]{REVERSEresidenceK0K2K4F0-2.pdf}
\caption{Waiting time distributions, $w(s)$ vs. translocation co-ordinate s, for $\kappa=$0, 2 and 4 respectively in Fig. a, b and c for $F_{0}=0.2$. The plots are for few intermediate apex angles to show the transition of behavior. }
\label{fig:resreversek0k2k4F02}
\end{figure*}
\end{comment}

\begin{figure*}
\includegraphics[width=\textwidth]{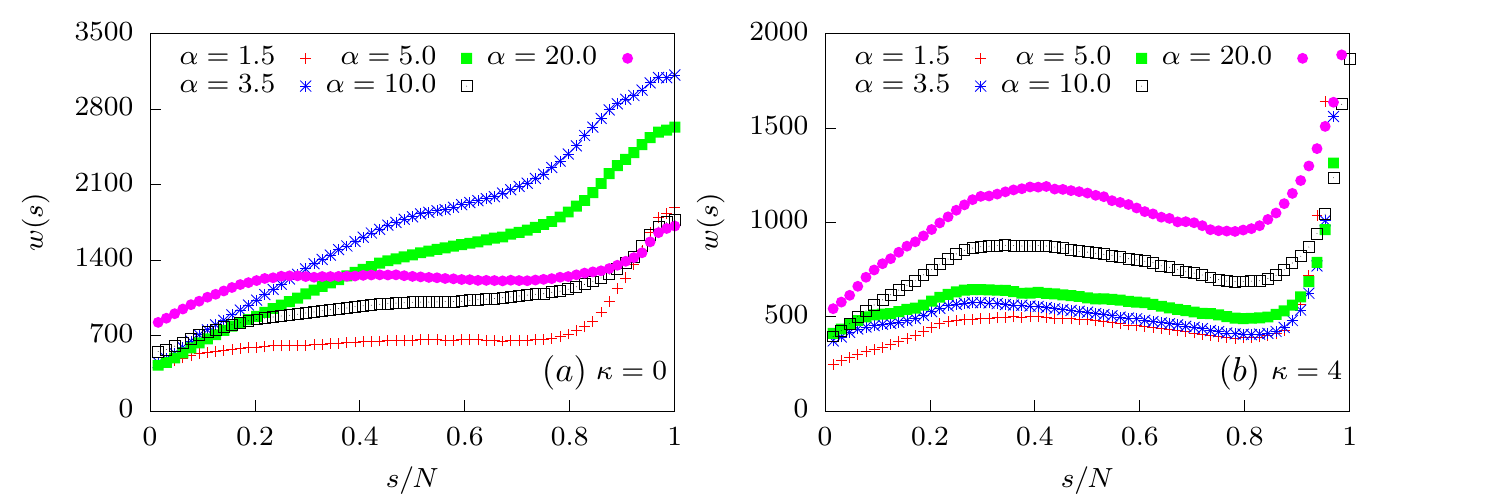}
\caption{Waiting time distributions, $w(s)$ vs. scaled translocation co-ordinate $s/N$ $(N=64)$, for $\kappa=$0, and 4 respectively in fig a, and b for $f_{0}=0.2$. The plots are for a few intermediate apex angles to show the transition in the $w(s)$ behavior. }
\label{fig:resreversek0k4F02}
\end{figure*}

%\subsection{Backward Translocation Dynamics:}
We first start with the case of a flexible polymer chain, $\kappa=0$. To see the effect of different pore geometries on the translocation process, we measure the average total translocation time  ($\tau$), one of the most fundamental properties of translocation dynamics. To study the effect of pore-polymer interaction at different stages of translocation, we measure the residence time $w(s)$ for each monomer $s$. Residence time is defined as the average time spent by monomer $s$ inside the channel during the translocation process. This is another important quantity that quantifies the effect of the pore geometry.
We find that for flexible polymer, the total translocation time ($\tau$) shows a non-monotonic behavior with the conical apex angle $\alpha$ (see Fig.~\ref{reversetransK0}). For a conical pore, the total translocation time, $\tau$, can be divided into two times, the passage time ($\tau_{p}$) and the escape time ($\tau_{e}$). The time taken for the first bead to reach the narrow exit, s.t., the pore is filled, is defined as the passage time, and the time from the passage phase to the time when the last bead exits the narrow end, is defined as escape time. Therefore, the total translocation time $\tau=\tau_{p}+\tau_{e}$. The pattern of $\tau$  vs. $\alpha$ is decided by the time components of the major contributions. In the case of a flexible polymer, we see from Fig.~\ref{reversetransK0}) that major contribution to the total translocation time $\tau$, is from the escape time component $\tau_{e}$ (empty circles). As evident from Fig.~\ref{forceplot}, the magnitude of the external force $f_{x}$ increases towards the narrower side of the pore for all $\alpha$'s. Thus, the filling of the pore becomes much easier with this external drive and hence the contribution of $\tau_{p}$ is much lesser in comparison to the escape time $\tau_e$. In Fig.~\ref{reversepercent}, the percentage contribution($p$) to the total $\tau$ is displayed for polymers of different rigidity. It can be seen that, irrespective of $\alpha$'s and $\kappa$'s, the major percentage contribution to $\tau$ is always from the escape time $\tau_{e}$. However, we see that, as the polymer becomes stiffer with the increase in $\kappa$, the relative $p$ contribution of $\tau_{e}$ increases in comparison to less flexible polymer chains and the relative contribution of $\tau_{p}$ decreases subsequently, Fig.~\ref{reversepercent}. Also, for $\alpha>5^{\circ}$, the passage $p$ gradually increases with increasing apex angle, $\alpha$. The increase in $p$ with increasing $\alpha$, itself signifies the increase in the cone area, which eventually increases the chance for the polymer of having coil-like conformations inside the cone while filling the pore. This increases the time for the first bead to reach the narrow opening of the conical channel. From the results, one can also anticipate getting the same $\tau$ vs. $\alpha$ behavior even if the simulation starts from the passage point. In the case of a flexible polymer, the percentage distribution shows non-monotonic behavior in both phases, passage as well as escape. This non-monotonic behaviour of $p$ contribution is clearly reflected in its total-translocation time $\tau$ vs. $\alpha$ plot as well, Fig.~\ref{reversetransK0}. Since the relation $\tau=\tau_{e}+\tau_{p}$ always holds, the passage and the escape distributions will always be a mirror image of each other.
\begin{figure}
\subfloat[]{\includegraphics[width=0.5\textwidth]{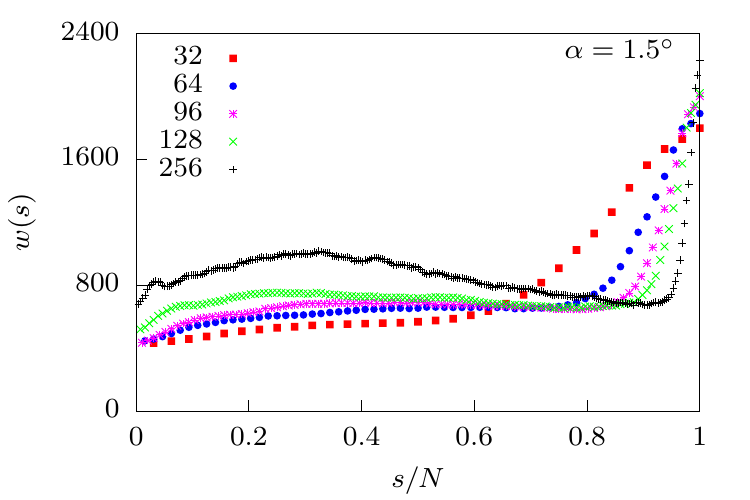}}\hspace{\fill}\quad
\subfloat[]{\includegraphics[width=0.5\textwidth]{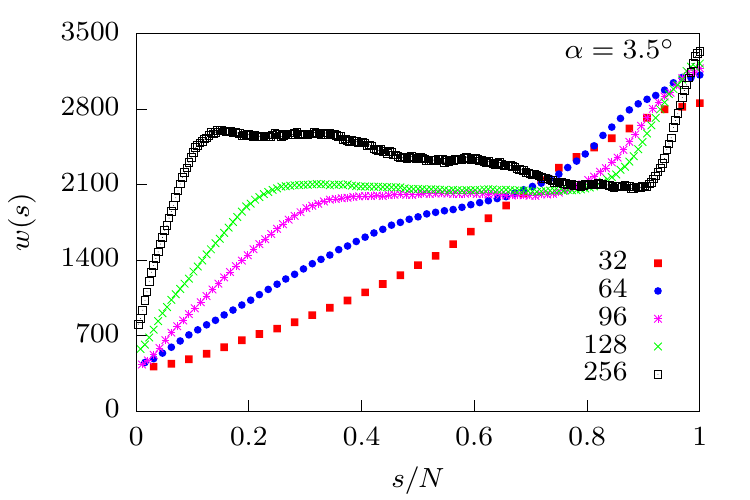}}\hspace{\fill}\quad
\subfloat[]{\includegraphics[width=0.5\textwidth]{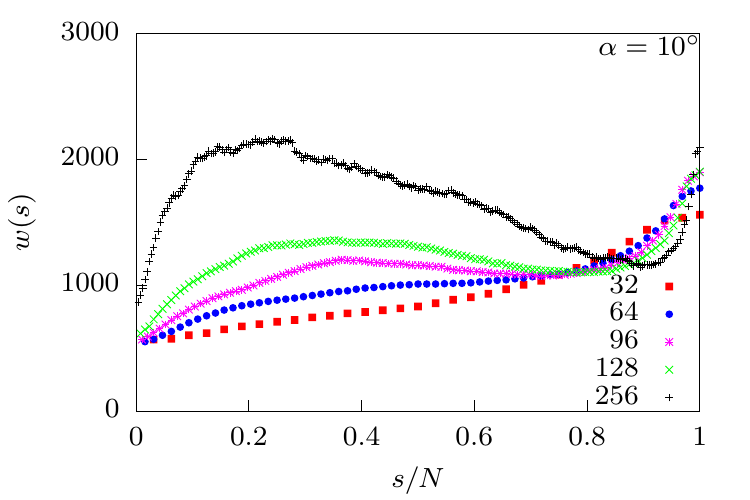}}
\caption{Waiting time distribution plots for flexible polymers with different polymer lengths ($L=$32, 64, 96, 128 and 256) for cone angle $\alpha=$ $1.5^{\circ}$, $3.5^{\circ}$ and $10^{\circ}$ in Fig. (a), (b) and (c) respectively. The graphs are plotted for $f_{0}=0.2$.}\label{fig:alphares}
\end{figure}

%And So, the requirement of the entropy force to drag the first bead to the narrow end increases, and with the decreasing magnitude of the external force at the wider conical areas, there is a delay in the pore-filling. Whereas, the relative escape-time $\%$ is seen to be gradually decreasing with increasing apex angle. In case of semi flexible polymers also, $\%$ contribution of $\tau_{e}>\tau_{p}$, fig.(\ref{fig:reversek0k2k4F02}d) where green is for $\kappa=2$ and blue is for $\kappa=4$.  The large difference in the two-time distributions also suggests that the polymer has to traverse less distance to fill the pore (or reach the passage point), which further gives a hint about the extent of the length of the pore.\\
%\textbf{Comparison with Forward translocation:}
To compare the results from different rigid polymers, we have plotted $\tau$ vs. $\alpha$  for four different rigid polymers for reverse translocation (filled circles), Fig.~\ref{forrevcompare} for lower force value, $f_{0}=0.2$. In the same figure, we have also plotted the results for the forward translocation case (filled squares) from \cite{sharma2022driven} so that we can compare them as well. For reverse case, the dependence of $\tau$ on $\alpha$ is non-monotonic in the case of a flexible polymer ($\kappa=0$). This behavior is similar to the case of forward translocation for a flexible polymer \cite{sharma2022driven}, where the translocation time vs. apex angle is non-monotonic at a lower force regime. The complexity of the problem makes it extremely difficult to know the underlying behaviors of the forward and backward translocation processes. The average translocation time $\tau$, however, shows a disparity between the forward and the reverse translocation processes in an intermediate cone apex angle ranging between $0.5^{\circ}$ and $9^{\circ}$. The range depends on the rigidity $\kappa$ of the translocating polymer and decreases with increasing the value of $\kappa$. For e.g., in the case of a flexible polymer, for $\alpha$ in the intermediate range $\alpha\in[0.5:9]$, the reverse translocation time is greater than the forward translocation time. For $\kappa=2$, the range shifts to $\alpha\in[0.5:8]$, and for $\kappa=4$, this range further shifts to $\alpha\in[0.5:6]$ Fig.~\ref{forrevcompare}(b-d). With the further increase in polymer rigidity, it is observed that the crossover time ($\tau_{c}$) keeps on shifting to a lower $\alpha$ value, which we call as crossover angle ($\alpha_{c}$) (see TABLE I). For $\alpha's$ greater than the crossover $\alpha_{c}$, forward translocation time is always greater than reverse translocation time. In this range of higher alpha values, $\alpha>\alpha_{c}$, the wider exit of the conical channel in forward case allows hairpin formation near the exit and hence it increases the escape time, while the single file escape of polymer segment from the narrow end makes the escape easier in the reverse case.
As seen in Fig.~\ref{forrevcompare}(a), $\tau_{reverse}\approx1.5\tau_{forward}$ for $\alpha\approx3^{\circ}$. This result is in complete agreement with the result of the ping-pong experiment performed with a fabricated conical pore with apex angle in the range ($2.6^{\circ}$ to $3.6^{\circ}$) \cite{asymmetric}. The translocation time in the experiment was found to be always large in the reverse flow, i.e from the wider opening to the narrow opening of the conical pore. From the geometry of the conical channel, we see that the wider y-extent for $3.5^{\circ}$ cone is apt enough to fit two beads simultaneously in a way that they are always in close proximity to both the attractive pore walls. The geometric construction and the entry of the flexible polymer from the wider end have the potential to form a block-like situation, where the polymer segment spends a lot of time inside the filled pore, either in a hairpin constriction or in a coiled form (see an embedded snapshot in fig.~\ref{reversetransK0}) and hence costs a lot of time to get out of the situation. One way to map the time delay is with the study of waiting time distributions ($w(s)$), which we will discuss in the next section. The bending energy associated with semi-flexible polymer segments does overcome the fluctuations inside the pore, resulting in the disappearance of the high differences in the translocation time between the above-mentioned intermediate alpha's. As can be seen in TABLE I, the crossover $\alpha_{c}$ keep shifting to a lower apex angle with the increase in rigidity of the polymer. So, one would expect that for a very rigid polymer chain, rigidity$\to\infty$, the translocation process will be independent of the direction of the channel and the $\tau$ vs. $\alpha$ curves for both forward and reverse cases will overlap each other.
\begin{figure*}
\subfloat[]{\includegraphics[width=0.49\textwidth]{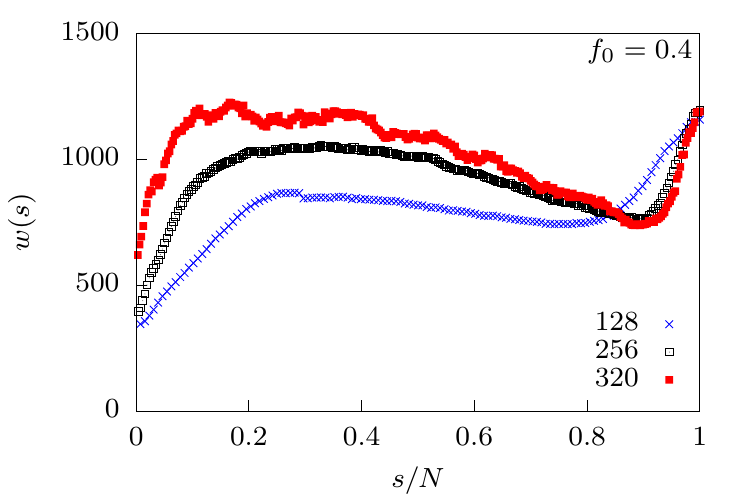}}\hspace{\fill}\quad
\subfloat[]{\includegraphics[width=0.49\textwidth]{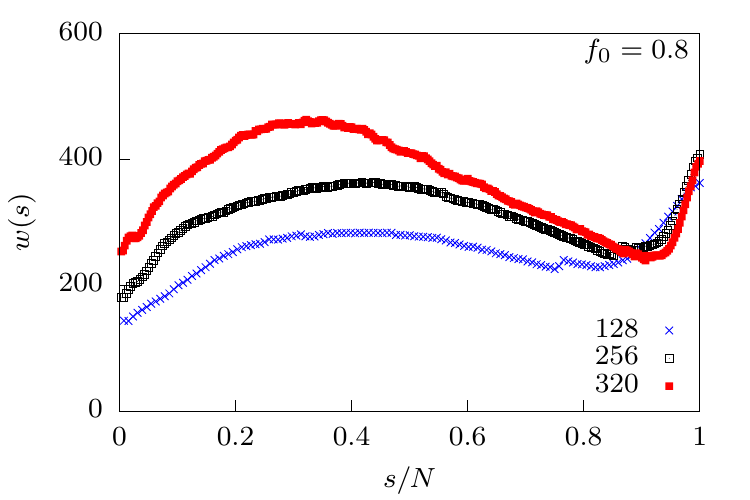}}\\
\subfloat[]{\includegraphics[width=0.49\textwidth]{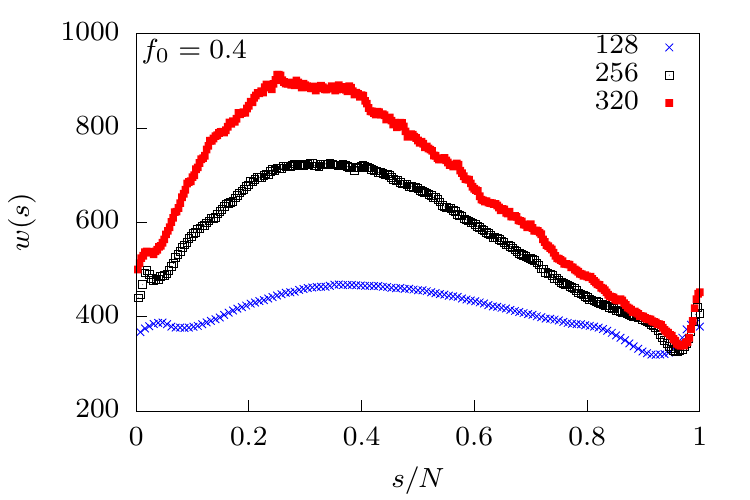}}\hspace{\fill}\quad
\subfloat[]{\includegraphics[width=0.49\textwidth]{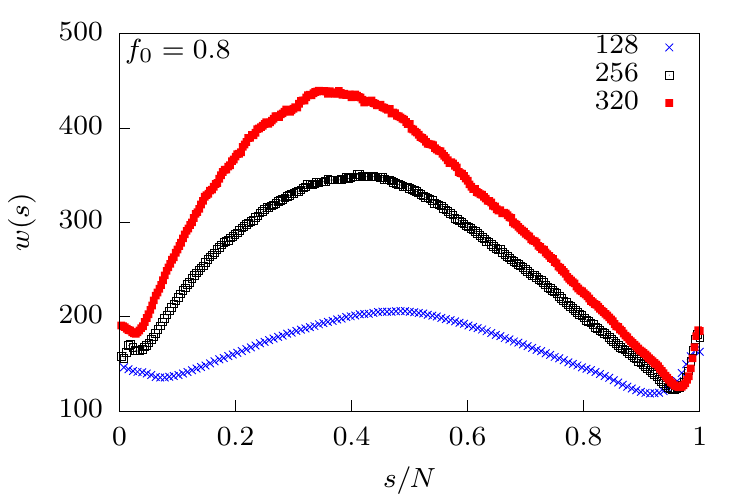}}
\caption{Comparing the effects of polymer lengths and force on $w(s)$ behaviour for reverse and forward translocation process of flexible polymers at $\alpha=3.5^{\circ}$. Plots (a-b) represents $w(s)$ for polymer lengths ($N=128,256,320$) for two force values ($f_{0}=0.4$ in (a), and 0.8 in (b))for the reverse translocation and plots (c-d) represents $w(s)$ for the forward translocation respectively.}\label{fig:alpha3-5res}
\end{figure*}

\begin{comment}
\begin{figure}
\subfloat[]{\includegraphics[width=0.45\textwidth]{forwardresitransA3-5K0F0-4.pdf}}\hspace{\fill}\quad
\subfloat[]{\includegraphics[width=0.45\textwidth]{forwardresitransA3-5K0F0-8.pdf}}
\caption{$w(s)$ for longer polymer chains in the forward translocation case.}\label{forward}
\end{figure}
\end{comment}

\begin{table}[]
\begin{tabular}{|c|c|c|}
\hline
$rigidity$  & crossover angle  &  $\tau_{c}(\approx)$)  \\ \vspace{-0.05cm}
($\kappa$)  & ($\alpha_{c}$) & \\  \hline
 0 &  $10^{\circ}$ & $3.2*10^{3}$   \\ \hline
2   & $9^{\circ}$ & $4.0*10^{3}$  \\ \hline
 4  & $8^{\circ}$  & $4.5*10^{3}$   \\ \hline
 8 &  $6^{\circ}$  & $7.5*10^{3}$ \\ \hline
 16 & $5^{\circ}$  & $1.4*10^{4}$ \\ \hline
 32 & $4.3^{\circ}$ & $3.5*10^{4}$    \\ \hline
64 &  $3.5^{\circ}$  & $10^{5}$   \\ \hline
\end{tabular}\caption{Values for the crossover angle ($\alpha_{c}$)  in the $\tau$ measure for the two processes: forward translocation and reverse translocation. With increasing rigidity ($\kappa$) of polymer, $\alpha_{c}$ shifts to lower alphas (col.2). Cross overtime ($\tau_{c}$), on the other hand, increases with an increase in rigidity of the polymer chain.}
\end{table}\label{table:table1}

%Also irrespective of rigidity, the total translocation time for $\alpha=0^{\circ}$ pore is always greater than other higher angles(as explained in the forward translocation paper). \\
%In $\alpha=0^{\circ}$, the confinement effect which is due to the lesser pore area in comparison to other higher $\alpha^{'}s$, plays a crucial role in confining the polymer segment during the whole transport process.\\

\begin{comment}
\begin{figure*}[]
\centering
\subfloat[]{\includegraphics[width=0.32\textwidth]{PhaseDiaF0-2K0K1K2A0-5A5reverse-1.png}}\hspace{\fill}\quad
\subfloat[]{\includegraphics[width=0.32\textwidth]{PhaseDiaF0-2K2K3K4reverse-1.png}}\hspace{\fill}\quad
\subfloat[]{\includegraphics[width=0.32\textwidth]{PhaseDiaF0-2K4K6K8A0A5reverse-1.png}}
\caption{}\label{phaseplot}
\end{figure*}
\end{comment}

\begin{comment}
\begin{figure}
\subfloat[]{\includegraphics[width=0.45\textwidth]{ReverseresitransA3-5K0F0-4.pdf}}\hspace{\fill}\quad
\subfloat[]{\includegraphics[width=0.45\textwidth]{ReverseresitransA3-5K0F0-8.pdf}}
\caption{Effect of increasing force on increasing chain length for $w(s)$ pattern of flexible chain at $\alpha=$ $3.5^{\circ}$. Fig.(a) is for $f_{0}=0.4$ and Fig.(b) is for $f_{0}=0.8$}\label{fig:alpha3-5res}
\end{figure}
\end{comment}

%\textbf{Residence plots:} 
The effect of the net pore polymer interaction on the translocation process can be studied from the waiting time distributions, $w(s)$. In Fig.~\ref{fig:resreversek0k4F02}(a-b) we have plotted the waiting time distribtions, $w(s)$, for two different polymer rigidities $\kappa=0$, and $\kappa=4$ respectively at lower force value, $f_{0}=0.2$. In the case of reverse translocation, the geometry of the conical pore allows single-file exit from the narrow end of the pore and this constraint can be used to track the overall escape process. Unlike the exit process, multiple beads can enter the pore simultaneously. This allows coiling and hairpin formation of the polymer segments near the entry zone which can block the pore and hence stalls the translocation process. The extent of stalling depends on the pore angle ($\alpha$) and the rigidity ($\kappa$) of the polymer. As, in case of a flexible polymer, the total translocation time for the $\alpha\in[2^{\circ}:6^{\circ}]$ differs a lot from rest of the non-zero $\alpha$ values, Fig.~\ref{forrevcompare}(a)(solid circles). This difference is also reflected in the waiting time distributions for flexible polymers, Fig.~\ref{fig:resreversek0k4F02}(a), where the $w(s)$ plots for the apex angle $3.5^{\circ}$(stars) and $5^{\circ}$(filled squares) exhibits very different behaviours than rest of the $\alpha$ values. For the apex angles $3.5^{\circ}$ and $5^{\circ}$, there are no retraction or plateau regions in $w(s)$ and it rises sharply as a function of scaled translocation coordinate($s/N$). This behaviour in the $w(s)$ is completely different from the $w(s)$ behaviour for other $\alpha$ values (see Fig.~\ref{fig:resreversek0k4F02}(a)), where $w(s)$ exhibits retraction features. It is interesting to note that, for all values of $\alpha$ and $\kappa$, Fig.~\ref{fig:resreversek0k4F02}(a-b), $w(s)$ always increases for the end monomers, \cite{sharma2022driven}. This reflects the fact that the escape from the narrow conical end is a single file translocation process. This result is in contrast with the forward case, where due to the multiple bead exit from the wider side of the conical pore, $w(s)$ decreases with $s/N$ during the pore emptying process, i.e. for the end monomers. For semiflexible polymer, Fig.~\ref{fig:resreversek0k4F02}(b), $w(s)$ shifts upward monotonically with apex angle for $\alpha\geq3.5^{\circ}$. This monotonic pattern, is also what we see in $\tau$ vs. $\alpha$ plots, Fig.~\ref{forrevcompare}(b-c).

\begin{comment}
\begin{figure}[H]
\includegraphics[height=2in]{reverseresitransA1-5K0F0-2.pdf}
\caption{Residence time plot for flexible polymer with different polymer lengths ($L=$32, 64, 96, 128 and 256) through conical pore ($\alpha=1.5^{\circ}$)}\label{fig:alpha1-5}
\end{figure}

\begin{figure}[H]
\includegraphics[height=2in]{ReverseresitransA3-5K0F0-2.pdf}
\caption{Residence time plot for flexible polymer with different polymer lengths through conical pore ($\alpha=3.5^{\circ}$)}\label{fig:Alpha3-5}
\end{figure}

\begin{figure}[H]
\includegraphics[height=2in]{reverseresitransA10K0F0-2.pdf}
\caption{Residence time plot for flexible polymer with different polymer lengths through conical pore ($\alpha=10^{\circ}$).}\label{fig:alpha10}
\end{figure}
\end{comment}

\begin{figure*}
\subfloat[]{\includegraphics[width=0.49\textwidth]{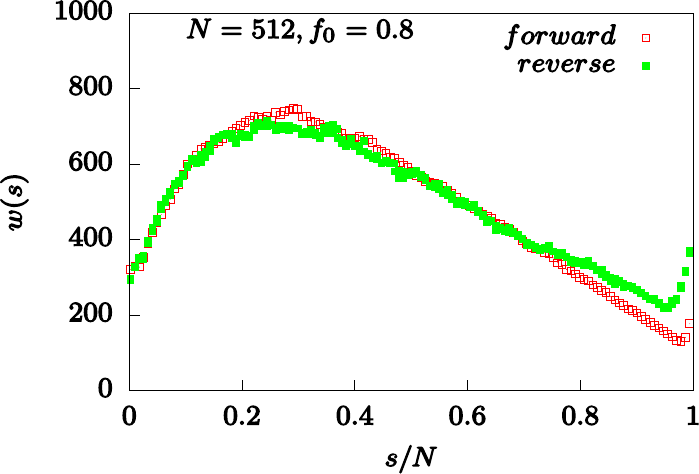}}
\subfloat[]{\includegraphics[width=0.49\textwidth]{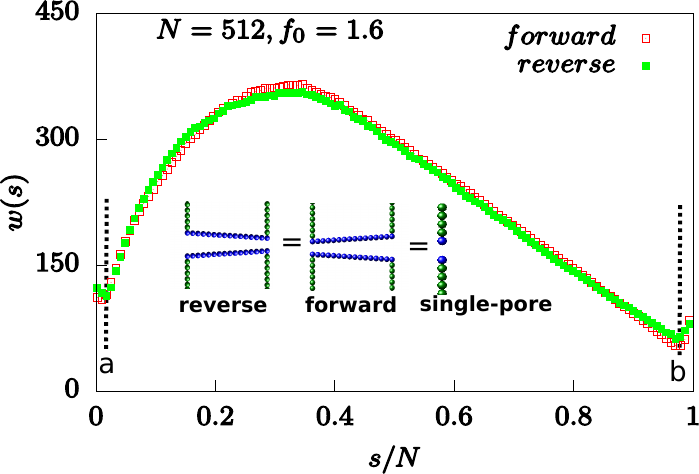}}
\caption{Plots showing the overlap of translocation properties in forward and reverse translocation process for polymer chain of length($N$) 512 through a conical pore ($\alpha=3.5^{\circ}$). The snapsorts embedded in fig.(b), shows the equivalence situation for longer chain length for the two conical pores (reverse and forward) with a single-pore between the point 'a' and 'b' of $w(s)$.}\label{poly512}
\end{figure*}

As mentioned above, for a flexible polymer chain of length 64, waiting time distributions is almost an increasing function of $s/N$ for $\alpha\in [3^{\circ}:5^{\circ}]$. %Here the plateau (tail retraction) pattern in $w(s)$ is absent which is otherwise present for other $\alpha$ values.
We have also studied the effect of polymer length($N$) on waiting time distributions ($w(s)$) for these $\alpha$ values. Figure ~\ref{fig:alphares}(b) shows the $w(s)$ plots for flexible polymers with chain lengths: 32, 64, 96, 128 and 256 for $f_{0}=0.2$. The effect of length on the translocation process can be witnessed with the appearance of plateau and tail retraction patterns in $w(s)$ with the increase in the polymer lengths. For smaller polymer lengths i.e, 32, $w(s)$ increases for subsequent transition coordinates ($s/N$), this shows the pure nature of stochastic behavior as no force balance scenario is experienced for any segment of polymers inside the pore. For $N=96$, the plateau begins at $s=19$ and persists till $s=48$. However as $N$ is increased from 96 to $N=$128(stars) and $N=$256 (square), fig.(\ref{fig:alphares}-b), the plateau in $w(s)$ transforms into tail retraction pattern. The emergence of tail retraction also shifts to a lower $s/N$ value with the increase in chain lengths. The plateaus are the result of a balanced force environment for the transporting beads. There are no net entropic force contributions from the segment on the cis side and the segment on the trans side. In this case, the pore beads are effectively only under the impact of a filled attractive pore and the driving force. To see the polymer length effect on $w(s)$ at lower and higher cone apex angles, we have plotted waiting time for various polymer lengths for $\alpha=1.5^{\circ}$ and $\alpha=10^{\circ}$ in Fig.~\ref{fig:alphares}(a) and Fig.~\ref{fig:alphares}(c), respectively. Here, we see that $w(s)$ is also an increasing function of $s/N$ for shorter polymer chain lengths (e.g. for $N=32,64$), but the rise in $w(s)$ is very slow in comparison to the abrupt rise in $w(s)$ for $\alpha=3.5^{\circ}$. In case of lower and higher cone apex angle, the tail retraction part is much enchaned for longer polymer chain lengths, compared to the retraction part of $w(s)$ for intermediate apex angle (i.e. for $\alpha=3.5^{\circ}$). Apart from this difference in $w(s)$, it is seen that for all apex angles, the translocation time $\tau$ increases linearly with the increase in the polymer length. We have also compared the $w(s)$ behaviour for longer polymer chains for reverse, Fig.~\ref{fig:alpha3-5res}(a-b), and forward translocation processes, Fig.~\ref{fig:alpha3-5res}(c-d). For both processes, we see that the tail retraction part becomes steeper with longer chains. However, for this range of $N$ values, we witness that in the case of reverse translocation, the propagation part (the initial rise in $w(s)$) is shorter than that for the forward case. But for very long polymer chain lengths (e.g. $N=512$), this disparity completely disappears and the translocation process appears to be direction-independent. In Fig~.\ref{poly512}, the overlapped waiting time distribution curve is plotted for both forward and reverse cases for $N=512$. With the increase in the magnitude of force, $f_{0}$, the retraction part of $w(s)$ is completely merged for both processes, Fig.~\ref{poly512}(b). Also, the overlapped region Fig.~\ref{poly512}(b) exhibits similar $w(s)$ behaviour for a polymer chain through a single-pore between the points 'a' and 'b'. The initial dip upto the point 'a', and the rise after the point 'b' in the graph restores the information of filling and exiting a extended and attractive extended pore. Hence, we state that, for the case of higher force and longer polymer chain lengths, the translocation dynamics through a filled extended pore is similar to the translocation dynamics of a polymer through a slit (\cite{adhikari2013driven}).

 \begin{figure*}

\subfloat[]{\includegraphics[width=0.321\textwidth]{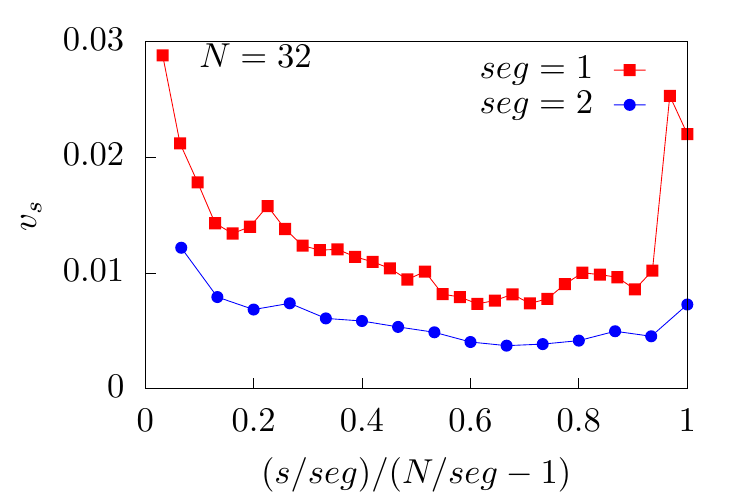}}\hspace{\fill}\quad
\subfloat[]{\includegraphics[width=0.321\textwidth]{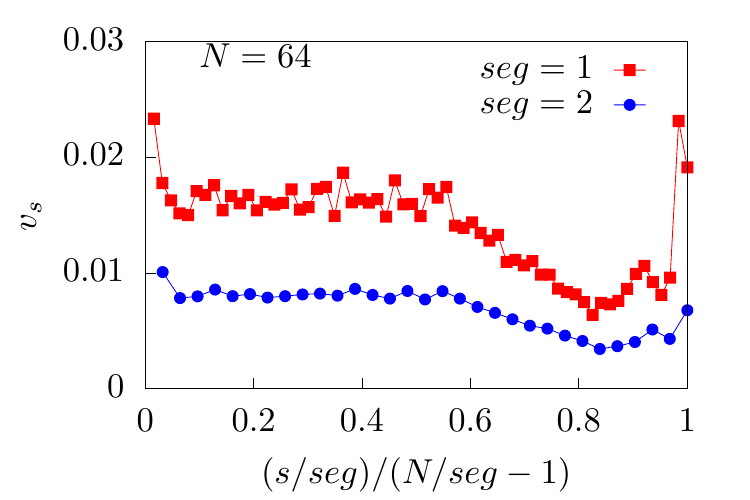}}\hspace{\fill}\quad
\subfloat[]{\includegraphics[width=0.321\textwidth]{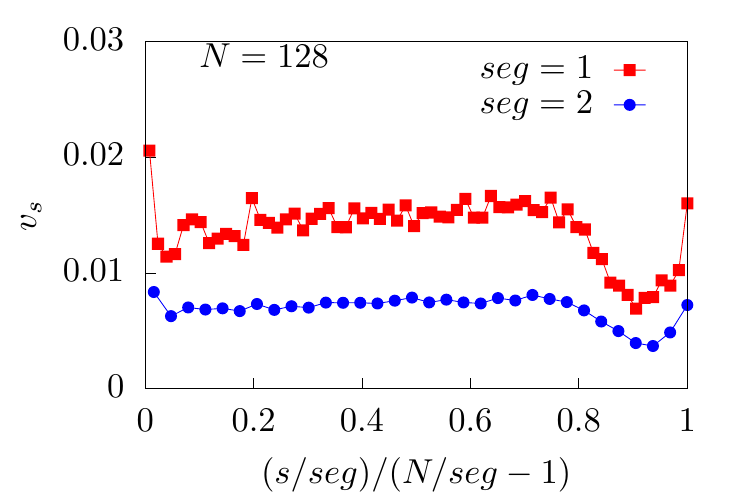}}
\caption{Plot showing the velocity of the exiting beads from the narrow end of the conical pore. It usually describes the time lapse before the next successful event (exit) of the following beads. In the x-axis of the plot, translocation coordinate $s$ is scaled by the total number of pairs formed by the seg-1 and seg-2 study. In the case of seg-1, the total number of tractable pairs will be 63, whereas for seg-2 total number of tractable pairs will be 31.  }\label{fig:velocitymarker1}
\end{figure*}

%\begin{figure}[H]
%\includegraphics[height=2in]{reverseresitransA3-5K0F0-2.pdf}
%\caption{Residence time plot for flexible polymer with different polymer lengths through conical pore ($\alpha=3.5^{\circ}$), inset plot shows the increasing behaviour of $\tau$ with polymer length(L).}\label{fig:alpha3-5}
%\end{figure}

\begin{comment}
\begin{figure*}
\includegraphics[height=4in]{velocitymarkerA3-5K0F0-2seperate.pdf}
\caption{Plot showing the velocity of the exiting beads from the narrow end of the conical pore. It usually describes the time lapse before the next successful event (exit) of the following beads. In the x-axis of the plot, translocation coordinate $s$ is scaled by the total number of pairs formed by the seg-1 and seg-2 study. In the case of seg-1, the total number of tractable pairs will be 63, whereas for seg-2 total number of tractable pairs will be 31.  }\label{fig:velocitymarker1}
\end{figure*}
\end{comment}

\begin{figure}
\includegraphics[width=0.5\textwidth]{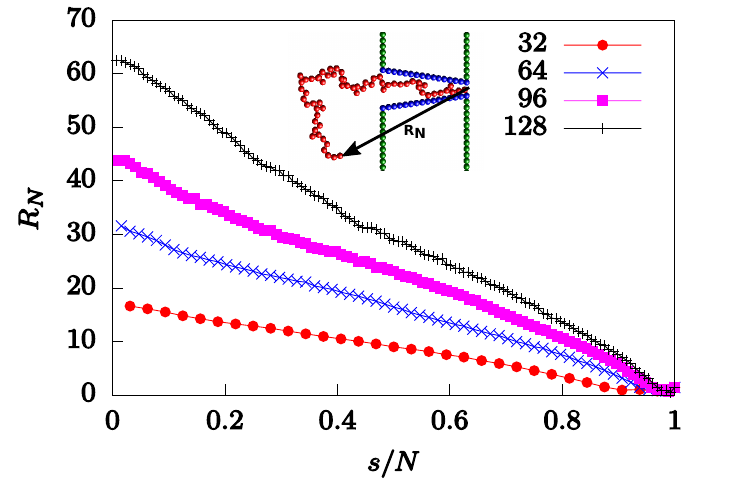}
\caption{ The distance of the last monomer $R_{N}$ of a flexible polymer chain through a reverse cone with $\alpha=3.5^{\circ}$ for $f_{0}=0.2$. N ranges from 32 to 128. The data was recorded once the pore is filled and the counter for s begins. The embedded figure represents the definition for $R_{N}$.}\label{enddistance}
\end{figure}

The changing behavior of $w(s)$ with increasing polymer lengths can also be studied from the velocity profile of the beads exiting from the pore to the trans side, \cite{chen2021dynamics}. For this, we tracked the velocity of successive beads through the narrow exit of the reverse pore, referred to as 1 segment (seg-1) velocity study, Fig.~\ref{fig:velocitymarker1}. And, similarly, the velocity for every alternative bead is referred to as 2 segments (seg-2) velocity study. For shorter polymer chain lengths, the escape velocity at the narrow end slows down during the translocation process, Fig.~\ref{fig:velocitymarker1}(a). For longer polymer chain length ($N=128$), Fig.~\ref{fig:velocitymarker1}(c), the escape velocity at the narrow end is constant for most of the translocation process and slows down as the pore starts emptying. For longer polymer chains, the attractive nature of the pore is seen during the pore emptying process whereas, for shorter chain lengths, the nature of the attractive pore effect dominates the overall translocation process. The constant-velocity region is more defined for 2-segment velocity profiles with very minimal fluctuations, in comparison to the 1-segment velocity profile. This implies a strong velocity correlation between exiting beads. It is also seen that irrespective of the chain lengths, the last bead is always pulled out at once. This exit is favored by the entropic force on the trans side. Figure ~\ref{enddistance} shows the variation in the end-to-end distance($R_{N}$) of the last bead of the chain from the narrow exit of the pore. The slope decreases linearly up to the value $(s-1)/N$, which implies that the last bead is still inside the pore. And once it crosses the pore from the narrow end, the slope increases, which implies that the polymer is successfully transported to the trans side of the pore. The $R_{N}$ behavior represents the fact that a flexible polymer always coils during the translocation dynamics, \cite{chen2021dynamics}. And, there is almost no possibility of attaining stretched polymer configuration for a lower force regime.

\section{CONCLUSION} \label{sec:discussion}
Based on our previous study(\cite{sharma2022driven}), we choose the force of interest at the narrow end, $f_{0}$, to be equal to 0.2 in this paper. To avoid the entry of beads from the wider end of the conical pores during the equilibration process, the pore was blocked during the whole equilibration process with additional repulsive beads and these additional beads are removed after the equilibration process. The reverse dynamics allows multiple bead entry of the equilibrated sample from the wider entrance of the conical pore, but the exit process is a single file translocation from the narrow end of the conical pore. For every successfully transported sample, the translocation time $\tau$ can be divided into two times: passage time and escape time, s.t. $\tau$ is always equal to the sum of the two times. It is seen that out of the two times, escape time has the major contribution to $\tau$. The combined effect of the cone asymmetry($\alpha$) and the flexibility of the polymer chains($\kappa$) plays an important role in determining $\tau$. For a flexible polymer, $\tau$ shows explicit non-monotonic behavior with the cone apex angle $\alpha$. While the non-monotonicity fades away with the rigid polymers. For higher apex angle, $\alpha\geq 7^{\circ}$, $\tau$ always increases with $\alpha's$, and for lower apex $\alpha\leq1.5^{\circ}$ ,$\tau$ always decreases with $\alpha's$. On comparing the forward and reverse translocation process, the maximum disparity in $\tau$ is seen for $\kappa=0$ for cone angle $\alpha\approx3.5^{\circ}$, where $\tau_{reverse}>>\tau_{forward}$. This finding is in agreement with \cite{bell2017asymmetric}, The delay in the polymer translocation in the reverse case is verified by the coil and hairpin formation of polymer segments near the wide pore entrance, which is not the scenario for the forward translocation case. %But for the intermediate $\alpha's$, $1.5^{\circ}<\alpha>7^{\circ}$, this statement is not always true. It is seen that for this range, the bending energy associated with $\kappa=2$ cannot withstand the fluctuations associated with a flexible chain. In this case, $\tau$ for $\kappa=0$ is greater than the $\tau$ for $\kappa=2$. Whereas for $\kappa=4$, bending energy does overpower the energy cost to escape, and hence transport time is always greater than flexible polymer, irrespective of $\alpha's$. Due to the wider region of entry, segments folding and coiling relatively slow down the polymer dynamics relative to the forward case, where entry is constrained to be a single file. The residence time plot of flexible polymer for extended pore is a true four-step process, but with an increase in apex angles or increase in rigidity of polymer chain, this four-step process switch to three step process. One common property in reverse dynamics is that $w(s)$ increases with translocation coordinate $s$, once the cis side is empty, and the exit to the trans side is governed by the pore-occupied beads.\\
We find that for a very rigid polymer chain or for a very long polymer chain length, the translocation dynamics through either side of a conical channel is a directional independent process. The translocation dynamics through a conical channel is dependent on the pore-polymer ratio. For a fixed pore length($L_{p}$) and pore angle($\alpha$), the behaviour of $w(s)$ changes with the increase polymer length, $N$. For a very long chain length($N\approx512$), the conical channel behaves as a single-pore. We have also observed that the transition of plateaus (tail retraction) with chain length can be studied from the one-segment velocity profile, where the transition behaviors in $v(s)$ can be mapped to that of their respective residence time ($w(s)$) behaviors. 

\section{Acknowledgement:}
I thank A. Chaudhuri and R. Kapri for their valuable suggestions and comments on the manuscript. I thank University Grants Commission (UGC), India for the financial support under the UGC-JRF/SRF scheme. I am thankful to IISER Mohali for providing a computational facility.
*Data can be availed upon reasonable request.

\bibliography{references}

\end{document}